\documentclass[conference]{IEEEtran}

\ifCLASSINFOpdf
   \usepackage[pdftex]{graphicx}
\else
   \usepackage[dvips]{graphicx}
\fi
\usepackage{cite}
\usepackage{amssymb}
\usepackage{amsmath}
\usepackage{booktabs}
\usepackage{mdwlist}

\begin{document}
%
\title{Efficient Privacy-Preserving Authentication Protocol for Vehicular Communications with Trustworthy}

\author{Hu Xiong$^{\dag}$$^{\ddag}$, Jianbin
Hu$^{\dag}$, Tao Yang$^{\dag}$, Wei Xin$^{\dag}$, Zhong
Chen$^{\dag}$
\\$^{\dag}$School of Electronics Engineering and Computer Science, \\Peking University, Beijing, P.R.China
\\$^{\ddag}$School of Computer Science and Engineering,
\\University of Electronic Science and Technology of China, Chengdu, P.R. China
\\Email: \{xionghu.uestc,hujianbin.pku\}@gmail.com, \{ytao,xinwei\}@pku.edu.cn, chen@ss.pku.edu.cn, }

\maketitle

\begin{abstract}

In this paper, we introduce an efficient and trustworthy conditional
privacy-preserving communication protocol for VANETs based on proxy
re-signature. The proposed protocol is characterized by the Trusted
Authority (TA) designating the Roadside Units (RSUs) to translate
signatures computed by the On-Board Units (OBUs) into one that are
valid with respect to TA's public key. In addition, the proposed
protocol offers both a \textit{priori} and a \textit{posteriori}
countermeasures: it can not only provide fast anonymous
authentication and privacy tracking, but guarantees message
trustworthiness for vehicle-to-vehicle (V2V) communications.
Furthermore, it reduces the communication overhead and offers fast
message authentication and, low storage requirements. We use
extensive analysis to demonstrate the merits of the proposed
protocol and to contrast it with previously proposed solutions.
\end{abstract}

\IEEEpeerreviewmaketitle
\section{Introduction}

Vehicular ad hoc networks (VANETs) are very likely to become the
most pervasive and applicable of mobile ad hoc networks (MANETs) in
this decade. Different from the traditional MANETs, VANET contains
not only mobile vehicles, but also stationary roadside
infrastructures. Equipped with communication devices, vehicles can
communicate with each other or with the roadside units (RSUs)
located at critical points of the road, such as intersections or
construction sites. Different from vehicles, RSUs usually have no
buffer constraint and can store a lot of information. According to
the Dedicated Short Range Communications (DSRC)\cite{DSRC}, each
vehicle equipped with OBU will broadcast routine traffic messages,
such as the position, current time, direction, speed,
acceleration/deceleration, and traffic events, etc. In this way,
drivers can get better awareness of the driving environment and take
early actions to the abnormal situation to improve the safety of
both vehicle drivers and passengers\cite{IVI}. However, before the
above attractive applications come into reality, the security and
privacy issues should be addressed. Otherwise, a VANET could be
subject to many security threats, which will lead to increasing
malicious attacks and service abuses. More precisely, an adversary
can either forge bogus messages to mislead other drivers or track
the locations of the intended vehicles. Therefore, how to secure
vehicle-to-vehicle communication in VANETs has been well-studied in
recent years
\cite{Raya2005,Raya2007,Sun2007,Lin2007,Lin2008a,Xiong2010,Lin2008b,Zhang2008a,Zhang2008b,Lu2008,Lu2009,Lu2010,Daza2009,Kounga2009,Wu2010,Wasef2010,Wang2008,Xi2007,Xi2008}.

Dealing with fraudulent messages in VANETs is a thorny issue due to
its inherent self-organization. The situation is further
deteriorated by the privacy requirements, i.e., the malicious
vehicles are anonymous and cannot be identified in case of dispute.
Countermeasures against fraudulent messages fall into two classes: a
\textit{posteriori} and a \textit{priori}.

With a \textit{posteriori} countermeasure, a trusted authority can
disclose the real identity of targeted OBU in case of a traffic
event dispute, even though the OBU itself is not traceable by the
public. In this way, punishment will be taken against vehicles who
have been proven to have originated fraudulent messages (e.g., the
violators will be excluded from the network). The existing
\textit{posteriori} solutions for VANETs can mainly be categorized
into following classes. The first one is based on a large number of
anonymous keys (denoted as LAB in the rest of this
paper)\cite{Raya2005,Raya2007}, the second one is based on a pure
group-oriented signature, such as group signature and ring signature
(denoted as GSB in the following)\cite{Sun2007,Lin2007,Xiong2010},
while the last one employs the roadside units (RSUs) to assist the
vehicle in authenticating messages (denoted as RSUB in the
following)\cite{Lu2008,Zhang2008a,Zhang2008b}. Though all of these
solutions can meet the conditional privacy requirement, they are in
vain against irrational attackers such as terrorists. Even for
rational attackers, damage has already occurred when punitive action
is taken.

A \textit{priori} countermeasure attempts to prevent the generation
of fraudulent messages. In this approach, a message is not
considered valid unless it has been endorsed by a number of vehicles
above a certain threshold. This approach is based on the assumption
that most users are honest, and therefore, they will not endorse any
message containing false data. To achieve this, messages received
must be distinguishable. The use of an honest majority to prevent
generation of fraudulent messages has previously been proposed in
\cite{Daza2009,Kounga2009,Wu2010}. However, although the underlying
assumption that there is a majority of honest vehicles in VANETs
generally holds, it cannot be excluded that a number of malicious
vehicles greater than or equal to the threshold are present in
specific locations. Furthermore, for convenience in implementation,
most of schemes assume that the threshold, i.e., the number of
honest vehicles in all cases, should be treated as a
\textit{one-size-fits-all} concept. However, we argue that threshold
is a \textit{scenario-specific} concept in the sense that different
scenario may have varying threshold requirements. Indeed, the
threshold should be adaptive according to the traffic density and
the message scope: A low density of vehicles calls for a lower
threshold, whereas a high density and a message relevant to all of
the traffic in a city require a sufficiently high threshold.

To address these issues, this paper proposes an efficient and
trustworthy conditional privacy preserving authentication protocol
for vehicle-to-vehicle communication based on proxy
re-signature\cite{Libert2008}. Compared to previous
message-authentication
schemes\cite{Raya2005,Raya2007,Sun2007,Lin2007,Lin2008a,Xiong2010,Lin2008b,Zhang2008a,Zhang2008b,Lu2008,Lu2009,Lu2010,Daza2009,Kounga2009,Wu2010,Wasef2010,Wang2008,Xi2007,Xi2008},
our scheme (which we dub PRSB) has the following unparalleled
features that, we believe, make it an excellent candidate for the
future VANETs:
\begin{itemize*}
\item \textit{Achieving both priori and posteriori
countermeasures}: Using the proxy re-signature to secure the
vehicle-to-vehicle communication, the RSUs can be allowed to
transform an OBU's signature into a TA's signature on the same
message. This conceals the unique identity of the OBU to prevent
information leakage to the malicious adversary, while still allowing
for internal auditing by the RSUs. Furthermore, the RSUs can
distinguish by itself whether the message was signed by the same
cheating vehicle multiple times or by multiple honest vehicles. By
this way, our scheme enables the RSUs only transform the messages
endorsed by a number of vehicles greater than or equal to a
threshold, and the vehicles endorsing cheating messages can later be
traced. We also note that a recent proposal in \cite{Wu2010} also
achieves both \textit{priori} and \textit{posteriori}
countermeasures by drawing on the linkable group signature.
\item \textit{Efficiency}:
Different from GSB protocols \cite{Sun2007,Lin2007,Wu2010}, the
proposed protocol can efficiently deal with a growing revocation
list and does not rely on updating the group public key and private
key at all unrevoked vehicles. Furthermore, our protocol does not
rely on a large storage space at each vehicle. Clearly, since the
OBU only need to generate the \textit{general} signature instead of
the \textit{anonymous} signature, the OBU communication and
computation overhead will be reduced at a fairly large scale.
\item \textit{Threshold-adaptivity}: The threshold in our
proposal can be adaptive according to the traffic context, unlike
most previous schemes in which the threshold has to be preset during
the stage of system initialization. This feature enables our
proposal to be deployed in complicated traffic scenarios.
\end{itemize*}

The remainder of this paper is organized as follows. Section
\ref{sec2} presents background information related to vehicular
network design and operation and surveys additional related work.
Section \ref{sec3} presents the problem formulation, system
architecture, and design objectives as well as the key cryptographic
techniques our solution is based on: bilinear maps and proxy
re-signatures. Section \ref{sec4} details the proposed security
protocol, followed by the security analysis and the performance
analysis in Section \ref{sec5} and Section \ref{sec6}, respectively.
Section \ref{sec7} concludes this paper.

\section{Background and Related Work}
\label{sec2}

\subsection{System Model}
\label{secIIA}

Similar to previous work\cite{Lu2008,Lu2009,Zhang2008a,Zhang2008b},
the considered system includes three types of entities: the top
Trusted authority (TA), the immobile RSUs at the roadside, and the
moving vehicles equipped with on-board units (OBUs).

\begin{figure}[!t]
\centering
\includegraphics[scale=0.42]{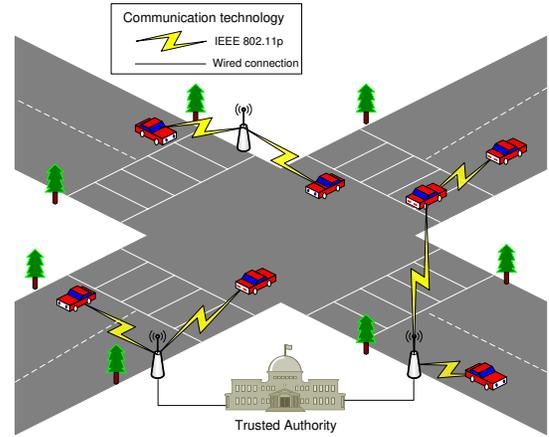}
\caption{Network Model} \label{fig1}
\end{figure}

\begin{itemize*}
  \item OBU: A vehicle needs to be registered to the TA with its public system parameters and corresponding private key
  before it joins the VANET.
  The secret information such as private keys to be used generates the need for a
  tamper-proof device in each vehicle. Similar to previous work we assume that access to this tamper-proof device
  is restricted to authorized parties.
  OBUs are mobile and moving most of the time. When the OBUs are on
  the road, they regularly broadcast routine safety
  messages, such as position, current time, direction, speed,
  traffic conditions, traffic events. The information system on each vehicle aggregates and processes these messages to enable drivers form a
  better awareness of their environment (Fig.
\ref{fig1}).
  The population of OBUs in the system could
  be up to billions (as, today, there are about a quarter of billion light vehicles in the US only).
  \item RSU: The RSUs are subordinated by the TA, which hold
storage units for storing information coming from the TA and the
OBUs. The main tasks of RSUs are (1) translating a OBU's signature
under the TA's public key on the same message, and (2) assisting the
TA to efficiently track the real OBU identity of any safety message.
Without the authorization of the TA, the RSUs will not disclose any
inner information. We remark that each RSU is physically secure and
cannot be compromised. Meanwhile, RSUs cannot generate signatures on
behalf of either the OBU or the TA. Different from the vehicles, we
assume that RSUs have neither computation and energy constraints nor
buffer size constraints. Due to the fact that there is no
computation and storage constraints at RSUs, RSUs can be able to
serve as the proxy to translate the signatures from OBUs.
  \item TA: The TA is in charge of the registration of all RSUs and OBUs each
vehicle is equipped with. The TA can reveal the real identity of
  a safety message sender by incorporating with its subordinate RSUs. To the end, the TA requires ample
  computation and storage capability, and the TA cannot be compromised and is fully trusted by
  all parties in the system.
\end{itemize*}

The network dynamics are characterized by quasi-permanent mobility,
high speed, and (in most cases) short connection times between
neighboring vehicles or between a vehicle and a roadside
infrastructure network access point. The assumed communication
protocol between neighboring OBUs or between an OBU and a RSU is 5.9
GHz Dedicated Short Range Communication (DSRC)\cite{DSRC} IEEE
802.11p.

\subsection{Related Work}

To achieve both message authentication and conditional anonymity,
Raya \emph{et al}.\cite{Raya2007,Raya2005} introduced the LAB
protocol. Their key idea is to install on each OBU a large number of
private keys and their corresponding anonymous certificates. To sign
each launched message, a vehicle randomly selects one of its
anonymous certificates and uses its corresponding private key. The
other vehicles use the public key of the sender enclosed with the
anonymous certificate to authenticate the source of the message.
These anonymous certificates are generated by employing the
pseudo-identities of the vehicles, instead of taking any real
identity information of the drivers. Each certificate has a short
life time to meet the drivers'privacy requirement. Although LAB
protocol can effectively meet the conditional privacy requirement,
it is inefficient and may meet a scalability bottleneck. The reason
is that a sufficient numbers of certificates must be issued to each
vehicle to maintain anonymity over a significant period of time. As
a result, the certificate database to be searched by the TA in order
to match a compromised certificate to its owner's identity is huge.
In addition, the protocols of \cite{Raya2007} are extended for
providing confidentiality in specific scenarios of VANET
implementations in \cite{Wang2008}. Subsequently, Lin \emph{et
al}.\cite{Lin2008b} developed the `time-efficient and secure
vehicular communication' scheme (TSVC) based on the TESLA (Timed
Efficient Stream Loss-tolerant Authentication)standard (RFC
4082)\cite{Perrig2002}. With TSVC, a vehicle first broadcasts a
commitment of hash chain to its neighbors and then uses the elements
of the hash chain to generate a message authentication code (MAC)
with which other neighbors can authenticate this vehicles' following
messages. Because of the fast speed of MAC verification, the
computation overhead of TSVC is reduced significantly. However, TSVC
also requires a huge set of anonymous public/private key pairs as
well as their corresponding public key certificates to be preloaded
in each vehicle. Furthermore, TSVC may not be robust when the
traffic becomes extremely dynamic as a vehicle should broadcast its
key chain commitment much more frequently.

Lin \emph{et al}.\cite{Lin2007,Lin2008a} proposed the GSB protocol,
based on the group signature\cite{Boneh2004}. With GSB, each vehicle
stores only a private key and a group public key. Messages are
signed using the group signature scheme without revealing any
identity information to the public. Thus privacy is preserved while
the trusted authority is able to track the identity of the sender.
However, the time for safety message verification grows linearly
with the number of revoked vehicles in the revocation list in the
entire network. Hence, each vehicle has to spend additional time on
safety message verification. Furthermore, when the number of revoked
vehicles in the revocation list is larger than some threshold, the
protocol requires every remaining vehicle to calculate a new private
key and group public key based on the exhaustive list of revoked
vehicles whenever a vehicle is revoked. Lin \emph{et
al}.\cite{Lin2007,Lin2008a}do not explore solutions to effectively
updated the system parameters for the participating to vehicles in a
timely, reliable and scalable fashion. This issue is not explored
and represents an important obstacle to the success of this scheme.
To address the scalability concern, Xiong \emph{et
al.}\cite{Xiong2010} proposed a spontaneous protocol based on the
revocable ring signature\cite{Liu2007}, which allows the vehicle to
generate the message without requiring online assistance from the
RSUs or the other vehicles. In this solution, the remaining vehicles
are not required to update their system parameters regardless of the
number of revoked vehicles. However, this protocol suffers larger
communication overhead than that of other protocols because the
length of ring signature depends on the size of the ring.

Recently, Zhang \emph{et al}.\cite{Zhang2008a,Zhang2008b} proposed a
novel RSU-aided message authentication scheme (RSUB), which makes
the RSUs responsible for verifying the authenticity of messages sent
from vehicles and for notifying the results back to vehicles.
Compared to the solutions previously mentioned, this scheme enables
lower computation and communication overheads for each vehicle.
Independently, Lu \emph{et al}. \cite{Lu2008} introduced an
efficient conditional privacy preservation protocol for VANETs based
on generating on-the-fly short-lived anonymous keys for the
communication between vehicles and RSUs. These keys enable fast
anonymous authentication and conditional privacy. Furthermore, Wasef
\emph{et al.}\cite{Wasef2010} proposed a RSUs-aided Distributed
Certificate Service (DCS) scheme along with a hierarchical authority
architecture. In this way, vehicles can update theirs pseudonymous
certificate sets from the RSUs. However, all of the above solutions
fall into the \textit{posteriori} countermeasures, which can only
exclude the rational attackers by punishing the malicious users
after the attack.

To reduce the damage to a bare minimum, the \textit{priori}
countermeasures have been proposed to prevent the generation of fake
messages. In this approach, a message is not considered valid unless
it has been endorsed by a number of vehicles above a certain
threshold. Most recently, Kounga \emph{et al.}\cite{Kounga2009}
proposed a solution that permits vehicles to verify the reliability
of information received from anonymous origins. In this solution,
each vehicle can generate the public/private key pairs by itself.
However, the assumption in this solution is very restricted in that
additional hardware is needed on the OBU. After that, a proposal is
also presented following the \textit{priori} protection paradigm
based on threshold signature by Daza \emph{et al.}\cite{Daza2009}.
Nevertheless, to obtain the anonymity, this protocol assumes that
the OBU installed on the vehicle can be removable and multi OBUs
could alternatively be used with the same vehicle (like several
cards can be used within a cell phone in the same time). Thus, this
assumption may enable malicious adversary to mount the so-called
Sybil attack: vehicles using different anonymous key pairs from
corresponding OBUs can sign multiple messages to pretend that these
messages were sent by different vehicles. Since multi OBUs can be
installed on the same vehicle, no one can find out whether all of
these signatures come from the same vehicle or not. After that, Wu
\emph{et al.}\cite{Wu2010} proposed a novel protocol based on
linkable group signature, which is equipped with both
\textit{priori} and \textit{posteriori} countermeasures. However,
they face the same adverse conditions in GSB protocol in which the
verification time grows linearly with the number of revoked vehicles
and every remaining vehicle need to update its private key and group
public key when the number of revoked vehicles is larger than some
threshold.

\section{Preliminaries}
\label{sec3}

\subsection{Objectives}

To avoid reinventing the wheel, we refer the readers to other
works\cite{Lin2007,Raya2005,Wu2010} for a full discussion of the
attacker model. In the context of this work, we focus on the
following security objectives.
%

\begin{enumerate*}
  \item \emph{Efficient anonymous authentication of safety messages}: The
  proposed scheme should provide an \textit{efficient} and
\textit{anonymous} message authentication mechanism. First, all
accepted messages should be
  delivered unaltered, and the origin of the messages should be authenticated to guard against impersonation attacks. Meanwhile,
  from the perspective of vehicle owners, it may not be acceptable to leak
  personal information, including identity and location, while
authenticating messages. Therefore, providing a
  secure yet anonymous message authentication is critical to the applicability of VANETs. Furthermore, the
  proposed scheme should have low overheads for safety message verification and storage at OBUs.
  \item \emph{Efficient tracking of the source of a disputed safety
  message}: An important and challenging issue in these conditions is enabling the TA to retrieve a vehicle's real identity from its pseudo identity.
  If this feature is not provided, anonymous authentication
  can only prevent an outside attack, but cannot deal with an inside
  one. Furthermore, the system should not only provide safety message traceability to prevent inside attacks, but also have reasonable overheads for the revealing the identity of a message sender.
    \item \emph{Threshold authentication}: A message is viewed as trustworthy
only after it has been endorsed by at least $n$ vehicles, where $n$
is a threshold. The threshold mechanism is a \textit{priori}
countermeasure that improves the confidence of other vehicles in a
message. In addition, the threshold in the proposed scheme should be
adaptive, that is to say, the sender can dynamically change the
threshold according to the traffic context and scenarios.
\end{enumerate*}

\subsection{Bilinear Maps}

Since bilinear maps\cite{Boneh2001} are the basis of our proposed
scheme, we briefly introduce them here.

Multiplicative cyclic groups $(\mathbb{G},\mathbb{G}_{T})$ of prime
order $q$ are called bilinear map groups if there is an efficiently
computable mapping $\hat{e}:\mathbb{G}\times \mathbb{G}\rightarrow
\mathbb{G}_{T}$ with the following properties:

\begin{enumerate}
    \item Bilinearity: For all $g,h\in \mathbb{G}$, and $a,b\in
\mathbb{Z}$, $\hat{e}(g^{a},h^{b})=\hat{e}(g,h)^{ab}$.\item
Non-degeneracy: $\hat{e}(g,h)\neq 1_{\mathbb{G}_{T}}$ whenever
$g,h\neq 1_{\mathbb{G}}$.
\end{enumerate}

Such an admissible bilinear map $\hat{e}$ can be constructed by the
modified Weil or Tate pairing on elliptic curves. For example, the
Tate pairing on MNT curves\cite{Miyaji2001} gives the efficient
implementation, and the representations of $\mathbb{G}$ can be
expressed in $161$ bits when the order $q$ is a $160$-bit prime. By
this construction, the discrete logarithm problem in $\mathbb{G}$
can reach $80$-bit security level.

\subsection{Proxy Re-Signature}

Proxy re-signature schemes, introduced by Blaze, Bleumer, and
Strauss\cite{Blaze1998}, and formalized later by Ateniese and
Hohenberger\cite{Ateniese2005}, allow a semi-trusted proxy to
transform a delegatee¡¯s signature into a delegator¡¯s signature on
the same message by using some additional information. Proxy
re-signature can be used to implement anonymizable signatures in
which outgoing messages are first signed by specific users. Before
releasing them to the outside world, a proxy translates signatures
into ones that verify under a system's public key so as to conceal
the original issuer's identity and the internal structure of the
organization. Recently, Libert et al.\cite{Libert2008} have
introduced the first \textit{multi-hop} \textit{unidirectional}
proxy re-signature scheme wherein the proxy can only translate
signatures in one direction and messages can be resigned a
polynomial number of times. We use this scheme as the basis for our
efficient and trustworthy conditional privacy-preservation protocol.

\section{Efficient and Trustworthy Vehicular Communications Scheme}
\label{sec4}

This section describes in detail our efficient and trustworthy
privacy-preserving protocol for VANET. TA, the delegator, will
designate the RSUs translating signatures computed from OBUs, the
delegatee, into one that is valid w.r.t. TA's public key by storing
the re-signature key at the RSUs. Upon receiving OBU's signatures,
the RSUs can validate them and re-sign the message using the
re-signature key. This message can be anonymously authenticated by
any vehicle participating in the system by verifying this signature
(the only information needed for verification is the TA's public
keys). By this way, proxy re-signatures can be used to conceal
identities of the OBU. Furthermore, RSUs could log which OBU signed
the message for solving the dispute, but keep that information
confidential to the public.

The notation used throughout this paper is listed in Table
\ref{tbl1}. The proposed security protocol is an extension of proxy
re-signature scheme \cite{Libert2008} in order to support
conditional anonymity authentication with trustworthy. Specifically,
the proposed security protocol contains four phases, which are
described in the following paragraphs.

\begin{table}[h]\caption{Notations}
\label{tbl1}
\begin{tabular}{ll}\toprule
Notations  &  Descriptions \\
\hline
TA:  &  \textbf{T}rusted \textbf{A}uthority \\
$V_{i}$:  &  The $i$th vehicle  \\
$RSU_j$:  &  an RSU works at location $L_j$\\
$\mathbb{G}$, $\mathbb{G}_{T}$: & two cyclic groups of
same order $q$\\
$g$: & The generator of $\mathbb{G}$\\
$RID_{i}:$ & The real identity of the vehicle $V_{i}$\\
$ID_{i}:$ & The pseudo-identity of the vehicle $V_{i}$\\
$M:$ & A message sent by the vehicle $V_{i}$\\
$x_{i}:$ & The private key of the vehicle $V_{i}$\\
$X_{i}=g^{x_{i}}$: & The corresponding public key of the vehicle $V_{i}$\\
$x_{RSU_{j}}:$ & The private key of the RSU $RSU_{j}$\\
$X_{RSU_{j}}=g^{x_{RSU_{j}}}$: & The corresponding public key of the RSU $RSU_{j}$\\
$x_{TA}$: & The private key of the TA\\
$X_{TA}=g^{x_{TA}}$: & The corresponding public key of the TA\\
$\mathcal{H}_{1}(\cdot):$ & A hash function such as $\mathcal{H}_{1}:\{0,1\}^{*}\rightarrow \mathbb{Z}^{*}_{q}$\\
$\mathcal{H}_{2}(\cdot):$ & A hash function such as $\mathcal{H}_{2}:\{0,1\}^{*}\rightarrow \mathbb{G}$\\
$Enc_{\kappa}():$ & A secure symmetric encryption algorithm with\\
& secret key $\kappa$\\
$a\parallel b$ & String concatenation of $a$ and $b$\\
\bottomrule
\end{tabular}
\end{table}

\subsection{System Initialization}

Firstly, as described in section \ref{secIIA}, we assume each
vehicle is equipped with a tamper-proof device, which is secure
against any compromise attempts in any circumstance. With the
tamper-proof device on vehicles, an adversary cannot extract any
data stored in the device including key material, data, and codes
\cite{Raya2005}. We assume that there is a Trusted Authority (TA)
which is in charge of registering the RSUs and the OBUs installed on
the vehicles. Prior to the network deployment, the TA sets up the
system parameters for each OBU and RSU as follows:

\begin{itemize*}
  \item Let $\mathbb{G}$, $\mathbb{G}_{T}$ be two cyclic groups of
same order $q$. Let $\hat{e}:\mathbb{G}\times \mathbb{G}\rightarrow
\mathbb{G}_{T}$ be a bilinear map.
  \item The TA first randomly chooses $x_{TA}\in_{R}\mathbb{Z}^{*}_{q}$ as its private key, and computes $X_{TA}=g^{x_{TA}}$ as its public
  key. The TA also chooses two secure cryptographic hash functions
  $\mathcal{H}_{1}:\{0,1\}^{*}\rightarrow \mathbb{Z}^{*}_{q}$ and $\mathcal{H}_{2}:\{0,1\}^{*}\rightarrow \mathbb{G}$, and a
  secure symmetric encryption algorithm $Enc_{\kappa}()$ with secret
  key $\kappa$.
     \item The TA generates public/private key pair for each subordinated $RSU_{j}$ works at location $L_j$ as
  follows:
  \begin{itemize}
    \item The TA randomly selects an integer
  $x_{RSU_{j}}\in_{R}\mathbb{Z}^{*}_{q}$ and computes
  $X_{RSU_{j}}=g^{x_{RSU_{j}}}$.
  \item The TA sends the
  public/private key pair to $RSU_{j}$ through a secure channel.
  \end{itemize}
  \item Each vehicle $V_{i}$ with real identity $RID_{i}$ generates its public/private key pair as
  follows:
  \begin{itemize}
    \item The vehicle $V_{i}$ first chooses $x_{i}\in_R\mathbb{Z}^{*}_{q}$ as its private key, and computes $X_{i}=g^{x_{i}}$ as its public
  key.
    \item  $V_{i}$ randomly selects an integer
  $t_{i}\in_{R}\mathbb{Z}^{*}_{q}$ to determine the verification
  information of $X_{i}$: $a_{i}=\mathcal{H}_{1}(g^{t_{i}}\parallel RID_{i})$ and $b_{i}=(t_{i}+x_{i}\cdot
  a_{i})$. Then  $V_{i}$ sends $\{X_{i},RID_i,a_{i},b_{i}\}$ to TA.
    \item After receiving $\{X_{i},RID_i,a_{i},b_{i}\}$, TA checks whether the following equation holds:
  $$a_{i}\stackrel{?}{=}\mathcal{H}_{1}((g^{b_{i}}X_{i}^{-a_{i}})\parallel RID_{i})$$
  If it holds, then $\{X_{i},RID_i\}$ is identified as the valid public key and identity. Otherwise, it will be
  rejected. After that, the TA stores the
  $(X_{i},RID_{i})$ in its records.
  \item In the end, TA generates the re-signature key $R_{i}=X_{i}^{1/x_{TA}}=g^{x_{i}/x_{TA}}$ which allows turning signatures
from vehicle $V_{i}$ into signatures from TA, and sends the item
$(R_{i},X_i)$ to all RSUs through a secure channel.
  \end{itemize}
  \item Each vehicle is preloaded with the public parameters $\{\mathbb{G}, \mathbb{G}_{T}, q,
  X_{TA}, \mathcal{H}, Enc_{\kappa}()\}$. In addition, the tamper-proof device of each vehicle is
  preloaded with its private/public key pairs $(x_{i},X_{i})$ and corresponding anonymous certificates
  (these certificates are generated by taking the vehicle's pseudo-identity $ID_{i}$).
\end{itemize*}

\subsection{OBU Safety Message Generation}

The format of the safety messages sent by the OBU is defined in
Table \ref{tbl2}, which consists of five fields: message ID,
payload, timestamp, $RSU_j$'s public key and signature. The message
ID defines the message type, and the payload field may include the
information on the vehicle's position, direction, speed, traffic
events, event time, and so on. According to \cite{DoT2006}, the
payload of a safety message is 100 bytes. A timestamp is used to
prevent the message replay attack. The next field is $RSU_j$, the
public key of RSU which will translate signature computed from OBU.
The first four fields are signed by the vehicle, by which the
``signature" field can be derived. Table \ref{tbl2} specifies the
suggested length for each field.

\begin{table}\centering\caption{Message Format for OBU}
\label{tbl2}
\begin{tabular}{|c|c|c|c|c|c|c|c|}
  \hline
  Message ID & Payload & Timestamp & $RSU_j$'s Public Key & Signature
  \\\hline
  2 & 100 bytes & 4 bytes & 20 bytes & 20 bytes\\
  \hline
\end{tabular}
\end{table}

\begin{figure}[!t]
\centering
\includegraphics[scale=0.42]{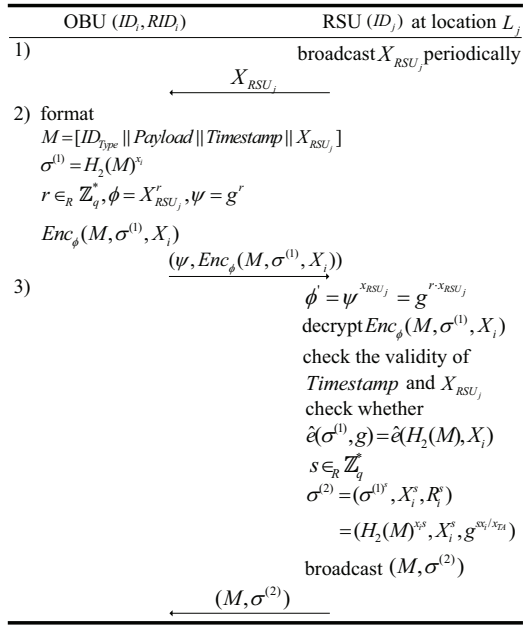}
\caption{ OBU Safety Message Generation} \label{alg}
\end{figure}

To endorse a message $M$, vehicle $V_{i}$ generates a signature on
the message, and then encrypts and sends it to $RSU_{j}$. After
receiving $n$ or more valid signatures from the vehicles, $RSU_{j}$
re-sign the message with the corresponding re-signature key and
broadcast the trustworthy signature. Fig. \ref{alg} shows the OBU
safety message generation, and the detailed protocol steps are
described as follows.

\begin{enumerate}
  \item $RSU_{j}$ broadcasts its public key $X_{RSU_{j}}$ periodically,
e.g., every 5 sec;
  \item $V_{i}$ computes signature $\sigma^{(1)}=\mathcal{H}_{2}(M)^{x_i}\in \mathbb{G}$ on message $M$, where
  $x_i$ is $V_{i}$'s secret key and $M$ is formatted as $[ID_{Type}\parallel Payload\parallel Timestamp\parallel X_{RSU_j}]$. Then, $V_{i}$ randomly chooses $r\in_{R}\mathbb{Z}^{*}_{q}$ and
  computes the shared secret key $\phi=X_{RSU_{j}}^{r}$ and the hint
  $\psi=g^{r}$.
  After that, $V_{i}$ sends the
  $(\psi,Enc_{\phi}(M, \sigma^{(1)}, X_i))$ to $RSU_{j}$;
  \item $RSU_{j}$ computes the shared secret key $\phi'=\psi^{x_{RSU_{j}}}=g^{r\cdot
  x_{RSU_{j}}}$ to decrypt the received message, and then looks up the newly updated revocation list from TA to check the validity of the public key $X_i$.
  After that, $RSU_{j}$ checks whether the
  signature $\sigma^{(1)}$ is valid as follows:
  $\hat{e}(\sigma^{(1)},g)\buildrel?\over=\hat{e}(X_i,\mathcal{H}_{2}(M))$.
 Then $RSU_{j}$ checks the validity of the RSU's
  public key and the freshness of timestamp embedded in the message.
  \item After receiving $n$ or more valid signatures from the vehicles on the same
  message $M$, $RSU_{j}$ search $(R_{i},X_i)$ according to $(M, \sigma^{(1)}, X_i)$ from its database. Then $RSU_{j}$ chooses randomly $s\in_{R}\mathbb{Z}^{*}_{q}$ and computes
  \begin{eqnarray*}
   \sigma^{(2)}=(\sigma_{0},\sigma_{1},\sigma_{2}) &=& (\sigma^{(1)^{s}},X_i^{s},R_{i}^{s}) \\
   &=& (\mathcal{H}_{2}(M)^{x_is},X_i^{s},g^{sx_{i}/x_{TA}})
  \end{eqnarray*} where
   $R_{i}$ have been preloaded along with $X_i$ in the $RSU_{j}$ during the
   initialization phase.
  Then $RSU_{j}$ stores the trace evidence table with item $(M,X_i)$ in its local
  database. In the end, TA
broadcast the trustworthy signature $(M,\sigma^{(2)})$ to all
vehicles among its coverage range.

  Note that the threshold $n$ can adaptively be changed according to the type of
message and various scenarios. For instance, if the message is an
alert about an emergency braking by the vehicle ahead, the threshold
can be set as low as 1. However, if the message is an announcement
that will affect many vehicles, the threshold can be set to be
appropriately high to improve the trustworthiness by also taking
into account the vehicle density among the RSU's communication
range. By this way, the signature $\sigma^{(1)}$ is turned into a
trustworthy signature $\sigma^{(2)}$ under TA's public key.
\end{enumerate}

\subsection{Message Verification}
\label{4-3}

Once a trustworthy message $\sigma^{(2)}$ is received, the receiving
vehicle performs signature verification by checking whether the
following conditions are true:

$\hat{e}(\sigma_{0},g)=\hat{e}(\mathcal{H}_{2}(M),\sigma_{1})$
  \qquad   \qquad $\hat{e}(\sigma_{1},g)=\hat{e}(\sigma_{2},X_{TA})$

This verification provides vehicles with the assurance that such a
signature can only have been computed if at least $n$ vehicles have
endorsed $M$.

\subsection{OBU fast tracing}
\label{secIVD}

If a vehicle produced a signature on the message $M$ and this
message was found to be fraudulent, a membership tracing operation
is started to determine the real identity of the signature
originator. In detail, the TA first position the RSU by extracting
the RSU's public key $X_{RSU_j}$ from the message
$[ID_{Type}\parallel Payload\parallel Timestamp\parallel
X_{RSU_j}]$. According to the TA's demand, the $RSU_j$ then
retrieves the public key of the source of the disputed safety
message $M$ by searching his trace evidence table with item
$(M,X_i)$ and returns $X_i$ to the TA, and then the TA recovers the
real identity from the returned public key.

\section{Security Analysis}
\label{sec5}

We analyze the security of the proposed scheme in terms of the
following four aspects: message authentication, user identity
privacy preservation, traceability by the TA, and threshold
authentication.
\begin{itemize*}
  \item \emph{Message authentication}. Message authentication is
  the basic security requirement in vehicular
  communications. In the proposed scheme, the signature $\sigma^{(1)}$ w.r.t public key $X_i$ can
  only be generated by the vehicle $V_i$, who holds the corresponding private key $x_i$.
  Without knowing the discrete logarithms $x_{i}$ of the public keys $X_{i}$,
  it is infeasible to forge a valid signature $\sigma^{(1)}$. If a signature $\sigma^{(1)}$ w.r.t public key $X_i$ passes the verification
procedure, it must be an intact fresh message generated by $V_i$.
This implies that the attacker cannot cheat RSU by forging a new
valid message, modifying an existing valid message, or replaying a
once valid but now expired message. Meanwhile, the signature
$\sigma^{(2)}$ can only be translated by the RSU from $\sigma^{(1)}$
by using the corresponding re-signature key $R_i$. Furthermore, the
RSU cannot generate the valid signature $\sigma^{(2)}$ on behalf of
$V_i$ using $R_i$. Thus, the adversary cannot forge the valid
signature $\sigma^{(2)}$ even when it only knows the corresponding
re-signature key $R_i$.
  \item \emph{Threshold authentication}. If a vehicle $V_i$ tends to cheat
RSU by endorsing the same message more than once, then the RSU can
easily link the multi signatures by comparing the public key $X_i$
along with the message. This kind of message can be either simply
discarded or sent to the TA to trace the cheating vehicle. Hence,
the Sybil attack can be avoided in our privacy-preserving scheme.
  \item \emph{Identity privacy preservation}. The message $M$ and the signature
  $\sigma^{(1)}$ with respect to public key $X_i$
  is only explored to $RSU_j$ and $V_i$ since the communication between $V_i$ and $RSU_j$ is
confidential. Finding the shared secret key $\phi$ from $\psi$ and
$X_{RSU_j}$ is an instance of the CDH problem: given $g$,
$\psi=g^{r}$, $X_{RSU_{j}}=g^{x_{RSU_{j}}}$, find $\phi=g^{r\cdot
  x_{RSU_{j}}}$. Thus, only the $RSU_j$ can link the $(X_i,\sigma^{(1)})$ to the corresponding message $M$. Given a
valid signature $\sigma^{(2)}$ of some message, it is
computationally difficult to identify the actual sending vehicle by
any vehicles in the system since the only information needed to
verify the correctness of signature $\sigma^{(2)}$ is TA's public
key $X_{TA}$.
  \item \emph{Traceability}. Given the disputed signature, only the corporation between TA and the $RSU_j$,
  can trace the real identity of a message sender using the OBU tracking procedure described in section \ref{secIVD}. Besides, the
  tracing process carried by the TA does not require any
  interaction with the message generator. Instead, the signature itself provides the authorship information to TA. Therefore, once a signature is in
  dispute, the TA has the ability to trace the disputed message, in
  which the traceability can be well satisfied.
\end{itemize*}

\section{Performance Evaluation}
\label{sec6}

This section evaluates the performance of the proposed scheme in
terms of storage requirements, and computational and communication
overheads.

\subsection{OBU Storage Overheads}

This subsection compares the OBU storage overhead of our protocol,
which we dub PRSB, with three previously proposed protocols:
LAB\cite{Raya2005,Raya2007,Lin2008b}, RSUB\cite{Lu2008} and
GSB\cite{Lin2007,Xiong2010,Wu2010}. In the LAB protocol, each OBU
stores not only its own $N_{okey}$ anonymous key pairs, but also all
the anonymous public keys and their certificates in the revocation
list (the notations adopted in the description are listed in Table
\ref{tbl3}). Let each key (with its certificate) occupy one storage
unit. If there are $m$ OBUs revoked, then the scale of revoked
anonymous public keys is $m\cdot N_{okey}$. Thus, the total storage
overhead in LAB protocol (denoted as $S_{LAB}$) is
$S_{LAB}=(m+1)N_{okey}$. Assuming that $N_{okey}=10^{4}$, we have
$S_{LAB}=(m+1)10^{4}$. In the GSB protocol, each OBU stores one
private key issued by the trusted party, and $m$ revoked public keys
in the revocation list. Let $S_{GSB}$ denotes the total storage unit
of GSB protocol. Thus, $S_{GSB}=m+1$. Both in the RSUB protocol
\cite{Lu2008} and our protocol, each OBU stores one public/private
key pair issued by the trusted party, and its anonymous certificate
issued by the RSU. Since the OBU does not need to store the
revocation list, the storage overhead in RSUB protocol is only two
units, denoted as $S_{RSUB}=S_{PRSB}=2$.

\begin{table}[h]\caption{Notations and rough scale}
\label{tbl3}\begin{center}
\begin{tabular}{lll}\toprule
&  Descriptions & Scale\\
\hline
$N_{obu}$  &  The number of OBUs in the system & $10^{7}$ \\
$N_{okey}$  &  The number of anonymous keys owned by one OBU & $10^{4}$ \\
$N_{rsu}$  &  The number of RSUs in the system & $10^{4}$ \\
$N_{rkey}$  &  The number of anonymous keys processed by one RSU & $10^{4}$ \\
\bottomrule
\end{tabular}\end{center}
\end{table}

\begin{figure}[!t]
\centering
\includegraphics[width=3.6in]{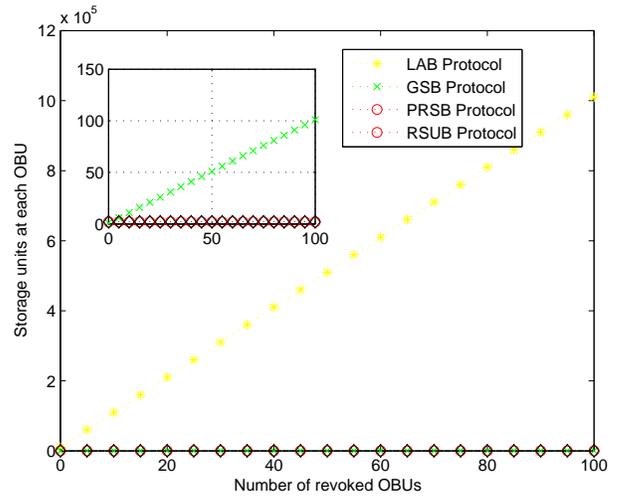}
\caption{OBU storage overhead for the protocols compared as a
function of the  number of revoked OBUs} \label{fig2}
\end{figure}

Fig.\ref{fig2} shows the storage units of LAB protocol, GSB
protocol, RSUB protocol and our protocol as $m$ increases. Observe
that the OBU storage overhead in LAB protocol linearly increases
with $m$, and is much larger than that in the other three protocols.
The storage overhead of GSB protocol is still small in spite of its
linear increase with $m$, while the storage overhead in the RSUB and
our protocol is the most efficient, which does not increase with
$m$.

\subsection{OBU Communication Overhead}

This section compares the communication overheads of the protocols
studied. We assume that all protocols generate a timestamp to
prevent replay attacks so we exclude the length of the timestamp in
this analysis.

For the LAB protocol, each message generates yields 181 bytes as the
additional overhead due to cryptographic operations, which includes
a certificate and an Elliptic Curve Digital Signature Algorithm
(ECDSA) signature\footnote{ECDSA signature scheme of
IEEE1609.2\cite{IEEE1609} is the current standard for VANETs, where
the length of a signature is 42 B.}. For the GSB$_{1}$
\cite{Lin2007}, GSB$_{2}$ \cite{Xiong2010} and GSB$_{3}$
\cite{Wu2010} protocol , each message generates $197$, $60n+60$ and
$133$ bytes as the additional overhead respectively, where $n$
represents the number of the public key pairs used to generate the
ring signature in \cite{Xiong2010}. For the RSUB protocols, the
additional communication overhead is $70/k+40+147$ bytes, where the
first term represents the communication overhead caused by
generating the short-term anonymous key, the second term represents
the length of the signature sent by the vehicle and the last term is
the length of the short time anonymous key and its corresponding
certificate which are reused across $k$ messages (as the RSUB
protocol regenerates the anonymous key only every $k$ messages). For
the proposed protocols, the additional communication overhead is
$2+20+20+20+20$ bytes, where the first term represents the
communication overhead caused by the message ID, the second term
represents the length of the $RSU_j$'s public key, the third term
represents the length of the signature sent by the vehicle, the
fourth term represents the vehicle's public key and the last term is
the length of the hint (as shown in Table \ref{tbl2}).

\begin{figure}[!t]
\centering
\includegraphics[width=3.6in]{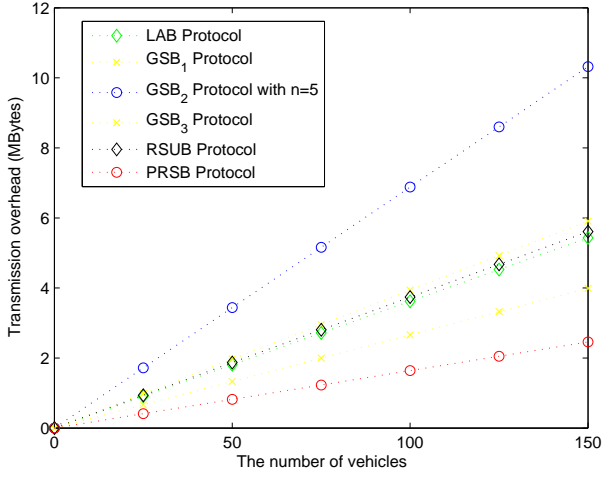}
\caption{Communication overhead versus traffic load.} \label{fig5}
\end{figure}

Fig. \ref{fig5} shows the relationship between the overall
communication overhead in 1 min and the traffic load within a
vehicle. Obviously, as the number of messages increases, the
transmission overhead increases linearly. Clearly, we can observe
that the proposed protocol has much lower communication overhead
than the other protocols.

\begin{table}[h]\caption{Comparison of communication overhead of three protocols}
\label{tbl5}\begin{center}
\begin{tabular}{lll}\toprule
Protocol &  Send a single message & Send $k$ messages\\
\hline
LAB  &  $181$ bytes & $181k$ bytes \\
GSB$_{1}$  &  $197$ bytes & $197k$ bytes \\
GSB$_{2}$  &  $60n+60$ bytes & $(60n+60)k$ bytes \\
GSB$_{3}$  &  $133$ bytes & $133k$ bytes \\
RSUB & $70/k+187$ bytes & $70+187k$ bytes \\
PRSB  &  $82$ bytes & $82k$ bytes \\
\bottomrule
\end{tabular}\end{center}
\end{table}

\subsection{OBU Computation Overhead}

This subsection compares the OBU computation overhead for the
proposed, RSUB and GSB protocols. Since the point multiplication in
$\mathbb{G}$ and pairing computations dominates each party's
computation overhead, we consider only these operations in the
following estimation. Table \ref{tbl4} gives the measured processing
time (in milliseconds) for an MNT curve of embedding degree $k=6$
and 160-bit $q$. The implementation was executed on an Intel pentium
IV 3.0 GHz machine.

\begin{table}[h]\caption{Notations and estimated execution time for cryptographic operations}
\label{tbl4}\begin{center}
\begin{tabular}{lll}\toprule
&  Descriptions & Execution Time\\
\hline
$T_{pmul}$  &  Time for one point multiplication in $\mathbb{G}$ & 0.6 ms \\
$T_{pair}$  &  Time for one pairing operation & 4.5 ms \\
\bottomrule
\end{tabular}\end{center}
\end{table}

In our proposed protocol, verifying a message, requires $4T_{pair}$
as shown in section \ref{4-3}. Let $T_{PRSB}$ be the required time
cost in our protocol, then we have:
$$T_{PRSB}=4T_{pair}=4\times 4.5=18 (ms)$$

In the RSUB protocol, to verify a message, it requires
$3T_{pair}+11T_{pmul}$. Let $T_{RSUB}$ be the required time cost in
RSUB's protocol, then we have:
$$T_{RSUB}=3T_{pair}+11T_{pmul}=3\times 4.5+11\times0.6=20.1 (ms)$$

In the GSB protocol \cite{Lin2007}, the time cost to verify a
message is related to the number of revoked OBUs in the revocation
list. Thus the required time is demonstrated as follows:
$$T_{GSB_{1}}=6T_{pmul}+(4+m)T_{pair}=6\times0.6+(4+m)\times4.5 (ms)$$

\begin{figure}[!t]
\centering
\includegraphics[width=3.6in]{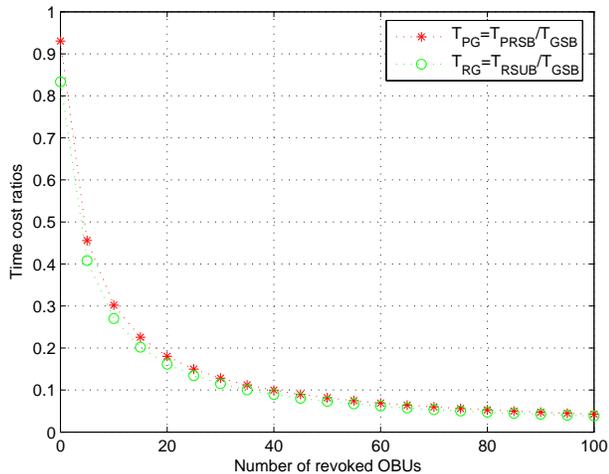}
\caption{Time efficiency ratio $T_{RG}=T_{RRSB}/T_{GSB}$ when
varying the number of revoked OBUs, $m$, from 1 to 100.}
\label{fig3}
\end{figure}

Let
$$T_{PG}=\frac{T_{PRSB}}{T_{GSB}}=\frac{4\times 4.5}{3.6+(4+m)\times 4.5}$$
$$T_{RG}=\frac{T_{RSUB}}{T_{GSB}}=\frac{3\times 4.5+11\times0.6}{3.6+(4+m)\times 4.5}$$
be the cost ratio between the PRSB and the GSB protocol, and between
the RSUB and the GSB protocol, respectively. Fig.\ref{fig3} plots
the time cost ratio $T_{PG}$ and $T_{RG}$ when $m$ OBUs are revoked,
as $m$ ranges from 1 to 100. We observe that both of the time cost
ratios decreases as $m$ increases, which demonstrates the much
better efficiency of our proposed protocol and RSUB protocol than
the GSB protocol especially when the revocation list is large. We
also observe that our proposed protocol is a little more efficient
than RSUB protocol.

%

\section{Summary}
\label{sec7}

We have presented an efficient conditional privacy preserving
protocol with trustworthy based on the proxy re-signature and aimed
for secure vehicular communications. We demonstrate that proposed
protocol is not only provides conditional privacy, a critical
requirement in VANETs, but also able to improve the confidence of
message receiver. By this way, our protocol achieves both
\textit{priori} and \textit{posteriori} countermeasures
simultaneously. Through extensive performance evaluation, we have
demonstrated that the proposed protocol can achieve much better
efficiency than previously reported counterparts in terms of the
number of keys stored at each vehicle, communication overhead and,
message verification.



\bibliographystyle{ieeetr}
\setlength{\baselineskip}{13pt}

\end{document}